\documentclass[
    ,final            % use final for the camera ready runs
  ]
  {aipproc}

\layoutstyle{8x11double}

\newcommand{\degrees}{\,^\circ}
\newcommand{\msun}{\,M_{\odot}}
\newcommand{\tsun}{T_{\odot}}

\newcommand{\psr}[1]
{\ifthenelse{\equal{#1}{0737}}{PSR J0737$-$3039}{\ifthenelse{\equal{#1}{1534}}{PSR B1534+12}{\ifthenelse{\equal{#1}{1913}}{PSR B1913+16}{\ifthenelse{\equal{#1}{1802}}{PSR J1802$-$2124}{\ifthenelse{\equal{#1}{1756}}{PSR J1756$-$2251}{\ifthenelse{\equal{#1}{1906}}{PSR J1906+0746}{\ifthenelse{\equal{#1}{1141}}{PSR J1141$-$6545}{\ifthenelse{\equal{#1}{2127}}{PSR B2127+11C}{\ifthenelse{\equal{#1}{0751}}{PSR J0751+1807}{\bf ???????}}}}}}}}}}

\newcommand{\lesssim}{\stackrel{<}{_\sim}}

\newcommand{\nodata}{\ldots}

\newcommand{\tcentre}{53000}

\begin{document}

\title{The double pulsar: evolutionary constraints from the system geometry}


\author{R.~D.~Ferdman}{
  address={Department of Physics and Astronomy, University of British Columbia, Vancouver BC, V6T 1Z1, Canada}
}

\author{I.~H.~Stairs}{
  address={Department of Physics and Astronomy, University of British Columbia, Vancouver BC, V6T 1Z1, Canada}
%{<common address for author2 and author3>}
}

\author{M.~Kramer}{
  address={University of Manchester, Jodrell Bank Observatory, Macclesfield, Cheshire, SK11 9DL, United Kingdom}
%  ,altaddress={<author1 address>} % additional visiting address
}

\author{R.~N.~Manchester}{
  address={Australia Telescope National Facility, CSIRO, P.O.~Box 76, Epping NSW 1710, Australia}
%  ,altaddress={<author1 address>} % additional visiting address
}

\author{A.~G.~Lyne}{
  address={University of Manchester, Jodrell Bank Observatory, Macclesfield, Cheshire, SK11 9DL, United Kingdom}
%  ,altaddress={<author1 address>} % additional visiting address
}

\author{R.~P.~Breton}{
  address={Department of Physics, McGill University, Montreal, QC H3A 2T8, Canada}
%  ,altaddress={<author1 address>} % additional visiting address
}

\author{M.~A.~McLaughlin}{
  address={Department of Physics, West Virginia University, Morgantown, WV 26505, USA}
%  ,altaddress={<author1 address>} % additional visiting address
}

\author{A.~Possenti}{
  address={INAF - Osservatorio Astronomico di Cagliari, Loc.~Poggio dei Pini, Strada 54, 09012 Capoterra (CA), Italy}
%  ,altaddress={<author1 address>} % additional visiting address
}

\author{M. Burgay}{
  address={INAF - Osservatorio Astronomico di Cagliari, Loc.~Poggio dei Pini, Strada 54, 09012 Capoterra (CA), Italy}
%  ,altaddress={<author1 address>} % additional visiting address
}

\begin{abstract}
The double pulsar system \psr{0737}A/B is a highly relativistic double neutron star (DNS) binary, with a $2.4$-hour orbital period.  The low mass of the second-formed NS, as well the low system eccentricity and proper motion, point to a different evolutionary scenario compared to other known DNS systems.  We describe analysis of the pulse profile shape over 6 years of observations, and present the resulting constraints on the system geometry.  We find the recycled pulsar in this system, PSR J0737$-$3039A, to have a low misalignment between its spin and orbital angular momentum axes, with a $68.3\%$ upper limit of $6.1\degrees$, assuming emission from both magnetic poles.  
This tight constraint supports the idea that the supernova that formed the second pulsar was relatively symmetric, possibly involving electron-capture onto an O-Ne-Mg core.
\end{abstract}

%%%%%%%%%%%%%%%%%%%%%%%%%%%%%%%%%%%%%%%%%%%%%%%%%%%%%%%%%%%%%%%%%%%
%%
%% The below \maketitle command inserts the actual front matter data.
%% It has to follow the above declarations.
%%
%%%%%%%%%%%%%%%%%%%%%%%%%%%

\maketitle

%%%%%%%%%%%%%%%%%%%%%%%%%%%%%%%%%%%%%%%%%%%%
%% MAINMATTER
%%
%%%%%%%%%%%%%%%%%%%%%%%%%%%%%%%%%%%%%%%%%%%%%%%%%%%%%%%%%%%%%%%%%%%%%%%%%%%%
%% Headings:
%%
%% The aipproc supports three heading levels, i.e., \section,
%%	\subsection, and \subsubsection.
%%%%%%%%%%%%%%%%%%%%%%%%%%%%%%%%%%%%%%%%%%%%%%%%%%%%%%%%%%%%%%%%%%%%%%%%%%%%
%% Cross-references:
%%
%% Page numbers (\pageref) and headings can NOT be referenced in the class,
%% since before being produced, no page numbers are determined.
%%
%% Tables, figures, and equeations can be referenced by using the LaTex
%% 	commands \label and \ref. For references to equation numbers, \eqref
%%	can be used, which will print "(1)" (while \ref will result in "1").
%%
%%%%%%%%%%%%%%%%%%%%%%%%%%%%%%%%%%%%%%%%%%%%%%%%%%%%%%%%%%%%%%%%%%%%%%%%%%%%
%% Lists: 
%%
%% Standard "itemize", "enumerate", etc. list environments are supported.
%%%%%%%%%%%%%%%%%%%%%%%%%%%%%%%%%%%%%%%%%%%%%%%%%%%%%%%%%%%%%%%%%%%%%%%%%%%%
%% Urls:
%%
%% \url{} command is provided for documenting URLs.
%%%%%%%%%%%%%%%%%%%%%%%%%%%%%%%%%%%%%%%%%%%%

\section{Introduction}

The \psr{0737}A/B system \citep{bdp+03,lbk+04} has provided pulsar astronomers and neutron-star theorists with an abundance of astrophysical phenomena to study in more detail, and with more precision, than ever before.  
Among these is the formation and evolution of binary pulsars, and in particular how this system, and perhaps others like it, have formed and evolved to arrive at their current configurations.  

We have performed analysis confirming that perhaps the channel of binary evolution undergone by this system is somewhat different than that of other double neutron star (DNS) systems for which orbital parameters and component masses have been measured.  The remainder of this paper will focus on describing the direct observational constraints on this evolution that we have gained through study of this system.

\section{Formation and evolution of the double pulsar system}
\label{sec:form_0737}

It is the evolution of the progenitor to the B pulsar that is believed to have caused the differences in current system properties between the \psr{0737}A/B system and, for example, the \psr{1913} and \psr{1534} DNS binaries as we now observe them \citep[e.g.,][]{pdl+05}.
The properties of the \psr{1913} and \psr{1534} systems suggest that the companion neutron stars (NS) formed via a core-collapse supernova.  In particular, this is supported by their large respective eccentricities, $e=0.617$ and $0.274$, and their high transverse velocities, $v_{\mathrm{tr}}=88$ and $107\,\mathrm{km}\,\mathrm{s}^{-1}$ \citep[][respectively]{tw89,sttw02}.

The observed parameters of the \psr{0737}A/B system---$e=0.088$ and $v_{\mathrm{tr}}=10\,$km\,s$^{-1}$ \citep{ksm+06}---appear to be 
%inconsistent with this scenario.  These are 
much lower than typically expected if the second supernova applied a large natal kick to the system, though are not by themselves conclusive evidence against a core-collapse supernova event having occurred.  However, the mass of the second-formed NS, \psr{0737}B, is also low, with $m_B=1.2489\pm0.0007\msun$.  Taken together, these properties present important clues that point to a different evolutionary path that formed the second NS.  
 It has been suggested that this may best be explained by the progenitor to pulsar B having gone through an electron-capture supernova \citep{pdl+05,std+06}.  This may avoid large supernova kicks due to the hypothesized short timescale over which this type of event proceeds, which is much shorter than that needed for instabilities that produce large kicks to develop.  \citet{pdl+05} have computed models of this scenario and found that the critical mass for collapse of an O-Ne-Mg core should range from 1.366 to 1.375$\msun$.

It has been noted by \citet{std+06} that the NS progenitor must have a minimum mass equal to that of the neutron star \emph{plus} the mass equivalent of the NS binding energy.  The binding energy of the NS is given by $E_{\mathrm{B}} \simeq 0.084(M_{\mathrm{NS}}/M_\odot)^2\msun$ \citep{ly89,lp01}, corresponding to $M_{\mathrm{B}}\sim 0.13\msun$ for \psr{0737}B, approximately equal to the difference between the B pulsar mass and the estimates made by \citet{pdl+05} of the pre-supernova mass.  If indeed this pulsar was formed from a collapsing O-Ne-Mg core, this would mean that almost no baryonic matter would have been lost in the process.  The remaining energy will have gone into changing the orbital properties of the system, and/or contributing to a kick at the time of the supernova.
The measured mass of \psr{0737}B, as well as the low transverse velocity of the system, thus present tantalizing evidence that the progenitor of the B pulsar may have ended its life in this manner.

In addition, it is expected from accretion theory that the spin axis of the A pulsar will have become aligned with the total angular momentum of the binary system (well-approximated by the orbital angular momentum) as it gained matter donated by the B pulsar progenitor.  If the supernova undergone by the B pulsar is close to symmetric, this alignment will not be disturbed \citep{plp+04}. By contrast, if there is a large kick to the system, the resulting misalignment will equal the angle between the orbital planes of the system before and after the supernova event \citep[e.g.,][]{wkk00}.  
Several studies \citep[e.g.,][]{dv04,ps05,wkf+06,std+06}, have examined the explosion of the B progenitor.  Using the timing-derived proper motion \citep{ksm+06}, \citet{std+06} predict a post-supernova misalignment angle $\delta$ for \psr{0737}A of $\lesssim 11\degrees$.  \citet{wkh05} predict compatible values under different kinematic and progenitor-mass assumptions.
%The study conducted by \citet{std+06} predicts, based on kinematics, a post-supernova misalignment angle $\delta$ for \psr{0737}A of $\lesssim 10\degrees$.  
If the above studies are correct, measuring a low spin-orbit misalignment angle for \psr{0737}A, in conjunction with the low system eccentricity and transverse velocity, would provide crucial evidence in explaining the observed properties of this system.

\section{Geodetic precession and long-term profile changes}

According to general relativity, the spin axis of a pulsar in a binary system will precess about the total angular momentum vector of the system.
This occurs at a rate given by \citep{dr74,bo75}:
\begin{equation}
  \label{eqn:geodetic}
  \Omega_{1}^{\mathrm{s}} = \left(\frac{2\pi}{P_b}\right)^{5/3} \tsun^{2/3}\frac{m_2(4m_1 + 3m_2)}{2(m_1 + m_2)^{4/3}} \frac{1}{1-e^2},
\end{equation}
where in this formulation, $m_1$ and $m_2$ are respectively the pulsar and companion masses, expressed in solar masses, $e$ is the orbital eccentricity, $P_b$ is the orbital period, and $\tsun= GM_{\odot}/c^3 = 4.925490947\,\mu\textrm{s}$ is the mass of the Sun expressed in units of time.

The effect of geodetic precession on the spin axis orientation of this pulsar serves to change our line of sight through the pulsar emission region over time.  
The extent to which these effects can be observed depends on the spin and orbital geometries of the pulsar system.  In particular, this is the case for the angle of misalignment $\delta$ between the pulsar spin and orbital angular momentum, which defines the opening angle of the cone that is swept out by the spin axis over the course of a precession period.  These geometric parameters can be modeled and determined through long-term analysis of the pulse profile shape.
It has been shown that for \psr{1913}, the relative amplitudes of the two major pulse components are changing significantly over time, as is the separation between these components \citep{tw89,kra98}.  It has also been demonstrated that the \psr{1534} profile displays a change in shape over time that is attributed to precession by an amount that is consistent with the general relativistic prediction \citep{sta04}.  

\psr{0737}A has a precession period of $\sim75$ years \citep{lbk+04}, much shorter than those of \psr{1913} and \psr{1534}.  The effects of geodetic precession on the \psr{0737}A pulse profile should thus be more readily observable on a shorter timescale than for \psr{1913} and \psr{1534}, provided that $\delta$ is non-negligible.

\section{A new  search for geodetic precession effects on PSR J0737$-$3039A}

\subsection{Observations: past and present}

Using a data set that spanned almost 3 years, \citet{mkp+05} found no evidence that the width or shape of the \psr{0737}A profile was changing with time. 
They argued that this may be attributed either to a small misalignment angle $\delta$ or to the pulsar being at or near a precession phase of $0\degrees$ or $180\degrees$. They assumed in their analysis that the pulsar emits from a single magnetic pole and so having $\delta\sim 0\degrees$ implies a broad beam structure that wraps around the neutron star, which seems rather unlikely.  On the other hand, having the pulsar at a special precession phase is statistically improbable.

To further investigate, we have extended the above analysis to include data taken between June 2005 and April 2007 with the Green Bank Telescope, using the Green Bank Astronomical Signal Processor (GASP) backend.  These data were obtained through biannual concentrated observing campaigns that spanned approximately 1-2 weeks each, and which typically contained 5-6 observing sessions that each lasted 6-8 hours.   These observations were taken exclusively using the 820-MHz receiver, and in general, used up to $16\times 4$\,MHz frequency channels.  These data have been flux-calibrated in each polarization using the signal from a noise diode source that is injected at the receiver.  We then calculated an average pulse profile for each observing epoch.

\begin{figure}[tp]
    \includegraphics[width=0.46\textwidth]{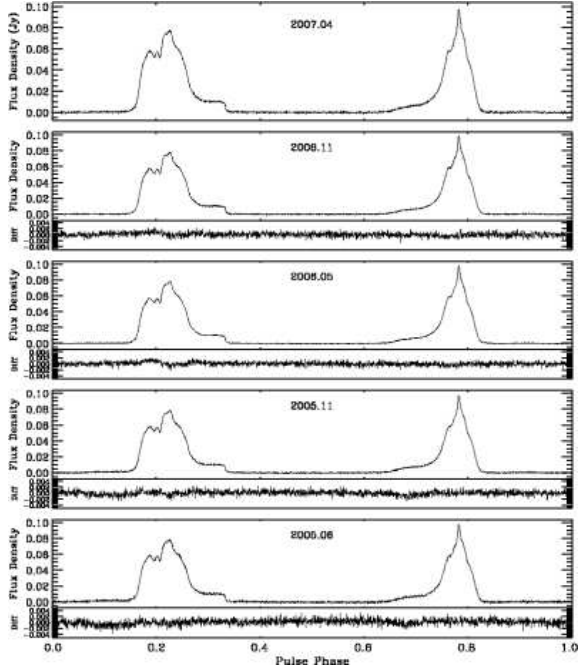}
    \caption{Pulse profiles and difference residuals over nearly two years of GBT observations.  High signal-to-noise, aligned profiles are shown from each of five concentrated observing campaigns.  The most recent of these, taken in April 2007, is shown in the top panel.  Each subsequent (shifted and scaled) profile and its difference from the April 2007 profile are shown and labelled in arbitrary flux units.  One can see the lack of significant pulse shape change over time. \label{fig:prof_diff}}
\end{figure}

Figure~\ref{fig:prof_diff} shows the aligned and subtracted profiles for each epoch.  While there seem to be subtle changes in the profile shape over the almost 2 years of observations, it is certainly not as apparent as one would expect if the spin-orbit misalignment is significantly non-zero.
The combined Parkes/GBT data set now spans nearly 6 years, or $\sim 7.5\%$ of the \psr{0737}A precession period; the hypothesis that the pulsar existing in a special phase of precession thus seems to be an unlikely explanation for the observed lack of profile variation. 

\subsection{Fitting for orbital geometry}

The dependence of the observed pulse width on the system geometry can be expressed by the following:
\begin{equation}\label{eqn:phi0}
  \cos{\Phi_0} = \frac{\cos\rho - \cos\zeta\cos\alpha}{\sin\zeta\sin\alpha},
\end{equation}
where $\Phi_0$ is half the full pulse width, the angle $\alpha$ is that between the spin and magnetic axes, $\zeta$ is the angle between the A pulsar spin axis and the observer's line of sight, and $\rho$ is the opening angle of the emission cone \citep{rl06a}.  In using this equation, we are assuming a circular emission beam (i.e.~no variation in $\rho$ with latitude on the pulsar surface).  In their work, \citet{mkp+05} investigated the effect of using a noncircular beam, and found that the results were only marginally affected.  We thus believe that a circular emission beam is a reasonable assumption for this analysis.  

As discussed earlier, geodetic precession causes the angle $\zeta$, describing our line of sight, to vary over time. It can be expressed as a function of time, as well as the misalignment angle $\delta$ and epoch of zero precession phase $T_1$, as follows \citep{dt92}:
\begin{equation}
\label{eqn:zeta_delta_t1}
  \cos{\zeta} = -\cos{\delta}\cos{i} + \sin{\delta}\sin{i}\cos{[\Omega_1^{\mathrm{spin}}(t - T_1)]},
\end{equation}
where $t$ is the observing epoch, and $i$ is the orbital inclination. In this analysis we assume that $\cos{i} > 0$.  The inclination $i=88.69_{-0.76}^{+0.50}\degrees$ reported by \citet{ksm+06} is sufficiently close to $90\degrees$ that the contribution from the first term in equation~\eqref{eqn:zeta_delta_t1} does not change by a significant amount by replacing $i$ by $180\degrees-i$.

If indeed the spin axis of the A pulsar is aligned, or nearly aligned, with that of the orbital angular momentum, then it follows that the spin axis should be nearly $90\degrees$ from the region of emission.  This would mean that this pulsar is an orthogonal rotator, presenting the possibility that we are viewing emission from both magnetic poles.  In investigating the geometry of the \psr{0737}A/B system, we performed analysis assuming both one- and two-pole emission models, where in the latter, each pulse component is treated as being emitted from separate magnetic poles.

\subsection{Geometric constraints}

The Parkes and GBT data were obtained at different observing frequencies, causing a systematic shift in measured pulse width due to pulse evolution with frequency.  In order to be able to use all the data in a single fit, we adjust the measured Parkes widths by subtracting from each of those measurements the difference between the weighted means of the GBT and Parkes data sets.  To obtain the errors on the adjusted Parkes data, the original Parkes data uncertainties are added in quadrature to that of the difference in the two means.
This subtraction is of course not physically correct, but it does allow the study of pulse shape evolution, since it retains the principle that there is zero pulse change within each data set.

\begin{figure}[tp]
  \begin{minipage}{\linewidth}
    \centering
    \includegraphics[width=0.49\textwidth]{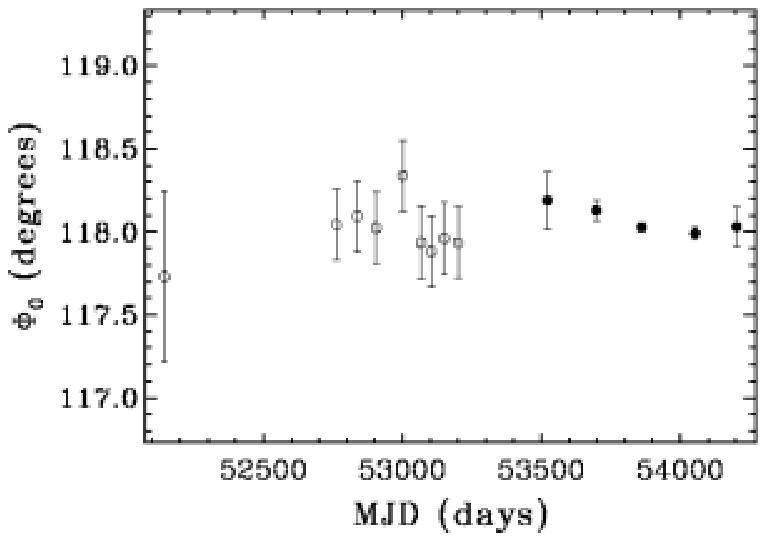}
    \begin{minipage}{0.5\linewidth}
      \centering
      \includegraphics[width=0.95\textwidth]{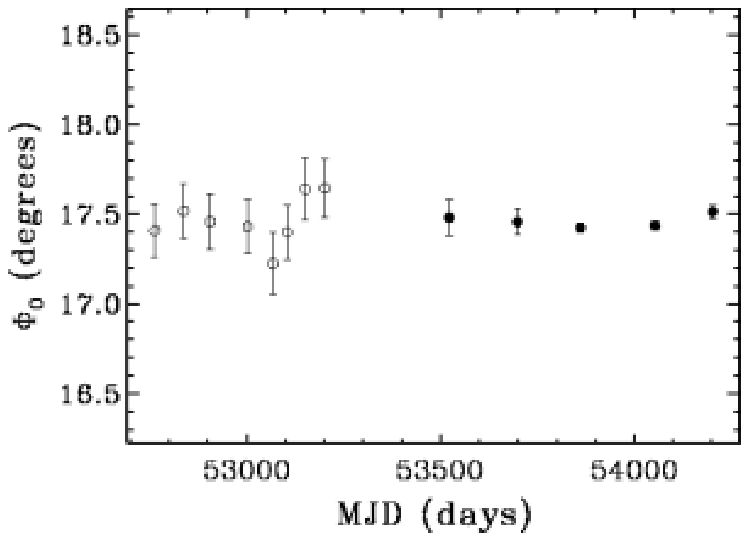}
    \end{minipage}\hfill
    \begin{minipage}{0.5\linewidth}
      \centering
      \includegraphics[width=0.95\textwidth]{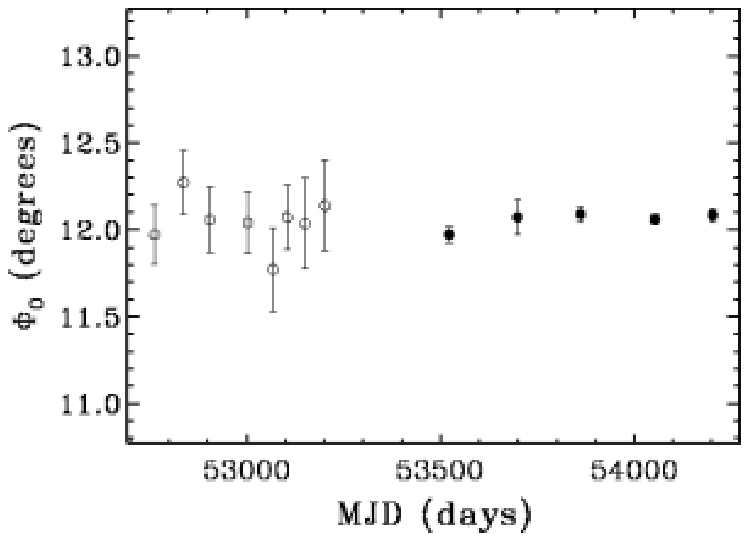}
    \end{minipage}
  \end{minipage} 
  \caption{Half-profile widths for PSR J0737$-$3039A as a function of MJD for Parkes filterbank data (open circles) and GASP data (filled circles).  The Parkes data set was adjusted by subtracting the difference of the (weighted) average widths of the Parkes and GBT data sets.  {\it Top}:  Widths measured at $10\%$ of the peak amplitude, assuming single-pole emission.  {\it Bottom}:  Widths measured at $25\%$ of the peak amplitude, for the first (left) and second (right) pulse components.\label{fig:0737_widths}}
\end{figure}

Figure~\ref{fig:0737_widths} shows the measured pulse widths assuming both one- and two-pole emission from \psr{0737}A.  For the former, we take our measurements at $10\%$ of the peak pulse height for consistency with the \citet{mkp+05} analysis.  For the latter, we re-determined the width of each pulse component separately, opting to use the $25\%$ peak flux width for these measurements.  Below this fractional pulse height, there are features in the profile that, when noise is added, negatively affect our measurements, making them unreliable. We chose to leave out the initial discovery data point from the two-pole analysis, since the extremely low signal-to-noise in that profile makes it difficult to obtain a reliable width measurement.  One can see that within each set of data there is no evidence for pulse shape change.

We performed fits on each set of width data using equation~\eqref{eqn:phi0}, probing the $\chi^2$-space over 2 sets of parameter grids.  The first pair of values we search are $\delta$ and $T_1$, with $0 < \delta < 180\degrees$ and with $T_1$ running over one precession period centred at MJD $\tcentre$.  The second grid runs over $\alpha$ and $\delta$, with $0\degrees \le \alpha \le 180\degrees$ and $0\degrees \le \delta \le 180\degrees$.  In all cases, we choose uniform prior distributions for these quantities within the respective constraints chosen, as we do not have information that would convince us to expect otherwise.  These fits were done using a Levenberg-Marquardt algorithm \citep[e.g.,][]{pftv86}, holding each pair of parameters fixed at each grid point, allowing the remaining two to vary as free parameters, arriving at a best-fit $\chi^2$ value correspoding to each $(\delta, T_1)$ and $(\alpha, \delta)$ combination, respectively.

For each grid, we arrived at a joint probability density function (PDF), from which we then calculated marginalized probability densities for each of the grid parameters.  The resulting constraints found for $\alpha$ and $\delta$ in the one- and two-pole cases are reported in Table~\ref{tab:alpha_delta_params_geo}.  In all cases we find that $\alpha$ is constrained to be close to $90\degrees$,  so that \psr{0737}A is indeed an orthogonal rotator.  The results shown for $\delta$ are those from the $(\delta, T_1)$ grid fit.  We have greater confidence in those upper limits on $\delta$, for the following reason:  There is no reason to prefer a specific epoch of zero precession over another.  This is a quantity that is determined by our line of sight to the pulsar, which we expect a priori to be uniformly distributed.  In the case of $\alpha$, however, we know less about the prior distribution, and so we choose a uniform prior for lack of a more compelling option.
 
In addition, the marginalized PDFs for $\delta$ between $0\degrees$ and $90\degrees$ are mirrored by those between $90\degrees$ and $180\degrees$, and it has been shown that having $\delta < 90\degrees$ is more physically likely, unless the supernova kick is extremely large, based on models of the misalignment angle in \psr{1913} \citep{bai88}.  We thus quote upper limits based on the PDFs for $\delta$ in the region below $90\degrees$.
We also find $T_1$ to be relatively unconstrained, which is expected for a low value of $\delta$.  It also supports the idea that it has become difficult to ascribe this lack of change to the pulsar being in a special precession phase, considering that our data spans approximately $7.5\%$ of the precession period for \psr{0737}A.
\begin{table}
\begin{tabular}{cccc|ccc|ccc}
\hline
% header
 & \tablehead{3}{c}{t}{One-pole model}  & \tablehead{6}{c}{t}{Two-pole model}\\
\tablehead{1}{c}{c}{Parameter} &   &   &   & \tablehead{3}{c}{b}{Pole 1} & \tablehead{3}{c}{b}{Pole 2} \\
\cline{2-10}
 & \tablehead{1}{c}{c}{Median}     & \tablehead{1}{c}{c}{68.3\%} & \tablehead{1}{c}{c}{95.4\%} & \tablehead{1}{c}{c}{Median} & \tablehead{1}{c}{c}{68.3\%}  & \tablehead{1}{c}{c}{95.4\%}  & \tablehead{1}{c}{c}{Median} & \tablehead{1}{c}{c}{68.3\%}  &   \tablehead{1}{c}{c}{95.4\%} \\
\hline%\hline
% data
%%% ALPHA %%%%
$\mathbf{\alpha}\ (^\circ)$  &  89.6    &  $70-99$      & $34-136$           &  89.6   &  $84-98$  &  $60-124$   & 91.1    & $87-101$  &  $69-121$  \\
\hline
%%% DELTA %%%%
$\mathbf{\delta}\ (^\circ)$  & \nodata  &   $< 15$        &  $< 37$          &\nodata &   $< 6.1$  &  $< 14$     & \nodata & $< 5.4$   &  $< 11$ \\
\hline
\end{tabular}
\caption{Summary of results for $\alpha$ and $\delta$ parameter estimation from long-term profile evolution analysis of PSR J0737$-$3039A.  Percentages reflect confidence intervals.}
\label{tab:alpha_delta_params_geo}
% footer
\end{table}

\paragraph{Aberration}

Based on our derived geometry for the \psr{0737}A/B system, we have predicted a very small latitudinal aberration signal as described by \citet{rl06a} (see also \citep{dt92}), which is expected to periodically distort the pulse profile over an orbital period.  Using the GBT data, we searched for, but found no significant evidence for the effects of aberration on the pulse profile of the A pulsar.

\subsubsection{One- or two-pole emission?}

For each of the one- and two-pole analyses, we calculated histograms of best-fit values found for $\rho$ at all grid points for the $(\alpha,\delta)$ and $(\delta,T_1)$ grid searches we performed.  In the single-pole emission case, we find that it is nearly impossible to avoid solutions with $\rho > 90\degrees$. It seems to be the case that if we constrain \psr{0737}A to be emitting from a single magnetic pole, that it would have to do so from a fan-shaped beam structure extending beyond a single hemisphere of the NS, possibly emitted from far in the magnetosphere.  This would mean that the beam centre is actually on the opposite side of the pulsar to where we expect, based on the profile symmetry.

In the two-pole analysis, we find that $\rho < 90\degrees$ is clearly favoured, with high occurrences of small values of $\rho$ within that range.  In this scenario, it is uncertain as to why the pulse peak separation is not $180\degrees$.
However, allowing emission to come from both magnetic poles in this pulsar thus avoids the need for exotically large beams, or else a reinterpretation of the the location of the emission beam centre.  

We thus favour a configuration in which \psr{0737}A is an orthogonal rotator with a spin axis that is nearly aligned with the orbital angular momentum of the system, and that the pulse profile that we observe is due to emission from both magnetic poles.  We then adopt $\alpha = 89.6^{+8.6}_{-5.8}\degrees$ , and the $68.3\%$ upper limit $\delta < 6.1\degrees$, based on the results from the first pulse component, which is the more conservative of the two.  

\section{Implications for evolution}

A low misalignment angle means that it is likely that a very small kick was imparted on the system due to the supernova of the pulsar B progenitor.  This agrees with the low-kick hypothesis that is favoured by measurements finding a low transverse velocity and eccentricity of the double pulsar, as well as the low mass of the B pulsar, all found from timing measurements of the double pulsar \citep{ksm+06}.  
It also lends credence to the studies performed by \citet{ps05}, \citet{wkf+06}, and \citet{std+06}, which favour a low-mass ($< 2\msun$) progenitor, and a low natal kick ($\lesssim 100\,$km\,s$^{-1}$) given the constraints of low space velocity.  In the case of the latter study, our constraints also agree with their estimate of the misalignment angle, which they predict to be $0.5 \le \delta \le 11\degrees$ ($95\%$ confidence).

This analysis thus supports a scenario in which the \psr{0737}B underwent a low mass-loss, relatively symmetric supernova event.
The prominent candidates for such an event are an electron-capture supernova, or a low-mass iron core collapse, and we therefore favour one of these scenarios as the one that produced the double pulsar system as we now observe it.  Discovery and observations of an increasing number of DNS binary systems will help to determine how prevalent this type of system is, and will provide further insight into this alternate channel of double neutron star formation.

\end{document}